# Do GRE scores help predict getting a physics Ph.D.?
# A comment on a paper by Miller et al.

Michael B. Weissman

*Department of Physics, University of Illinois at Urbana-Champaign*

*1110 West Green Street, Urbana, IL 61801-3080*

Abstract

A recent paper in Sci. Adv. by Miller et al. concludes that GREs do not help predict whether physics grad students will get Ph.D.s. The paper makes numerous elementary statistics errors, including introduction of unnecessary collider-like stratification bias, variance inflation by collinearity and range restriction, omission of needed data (some subsequently provided), a peculiar choice of null hypothesis on subgroups, blurring the distinction between failure to reject a null and accepting a null, and an extraordinary procedure for radically inflating confidence intervals in a figure. Release of results of simpler models, e.g. without unnecessary stratification, would fix some key problems. The paper exhibits exactly the sort of research techniques which we should be teaching students to avoid.

"The aim of science is not to open the door to infinite wisdom, but to set a limit to infinite error." — Bertolt Brecht

Introduction

A recent paper by Miller et al. (*1, 2*) argues, primarily with regard to the use of GRE scores, that "Typical Ph.D. admissions criteria limit access to underrepresented groups but fail to predict doctoral completion." They claim "The weight of evidence in this paper…indicates that lower than average scores on admissions exams do not imply a lower than average probability of earning a physics Ph.D." so that GREs are "metrics that do not predict Ph.D. completion." These are surprising conclusions to reach for a paper that frames results in terms of null-hypothesis p-value cutoffs (*3*) and that shows (see Table 2 of (*1*)) only one predictor for doctoral completion that can be used by admissions committees to select students and has p-value <0.01 in the overall sample studied: the GRE quantitative test (GRE-Q).

In this response I describe several improper statistical methods used in the article. My response is not intended to take a position on the complicated issue of desirable admissions criteria but only to defend minimal standards of research competence and transparency. I will not explore here whether other papers in the field make similar errors but will include some pedagogical material on elementary statistics. My core technical arguments have been published (*4*) along with a response from the original authors (*5*), which I will discuss in a later section of this paper.

Before evaluating the validity of reference (*1*), we need to clarify what question it is trying to answer. The statement "Our goal here was not to identify the best predictive model with the minimum number of parameters but rather to understand how all four commonly used admissions metrics (UGPA, GRE-Q, GRE-V, and GRE-P) and the most salient demographic information would contribute to a discussion of metrics and diversity by admissions committees" (*1*) does not help much. More succinctly, the main goal appears to be to estimate how much predictive power for degree completion would be lost by de-emphasizing or dropping the GRE components of the admissions criteria. More formally, one wishes to evaluate what effect a treatment (inclusion of GREs in admissions criteria) has on an outcome (Ph.D. rate).

Directly evaluating how well students admitted by different criteria would have done requires either a randomized trial in which similar programs would be randomly assigned to do GRE-aware or GRE-blind admissions (not feasible at the time of the study) or a comparison of non-randomly assigned programs using modern causal inference methods(*6*) to attempt to reduce systematic errors.  There seem not to have been enough GRE-blind programs to allow such an observational study. (*1*)  Instead, the strategy is to create an implicit model of what causes program completion, from which one can try to back out what the effect of dropping GREs would have been. Although that reasoning is not spelled out clearly, this plan would be reasonable if implemented properly.

The authors model Ph.D. attainment, a crude but convenient dichotomous proxy for broader ultimate goals such as scientific productivity, by a standard logistic regression, with the logit given by a multivariate linear regression on several predictors. Although use of this dichotomous outcome no doubt loses some important dynamic range compared to the outcomes of interest, it has the advantage of being easy to quantify without too much time delay. The multivariate form is justified as a way to give a better "basis for policy decisions" by avoiding "confounding" (*1*). Since confounding is a purely causal concept, these claims confirm that the results are intended to tell us what the causal effects of policy choices would be. Specifically, the model coefficients for the predictors, combined with the ranges of the predictors, are intended to tell us how much incremental predictive power would be lost by dropping each predictor, i.e. what the causal effects of that policy change would be on graduation rate. The predictors include percentile ranks of GRE scores (quantitative GRE-Q, verbal GRE-V, and physics GRE-P), undergraduate GPA, gender, ethnicity/race, U.S. vs. non-U.S. citizenship, year of matriculation, and one predictor that an admissions committee constrained by causality cannot use as an attribute to distinguish between applicants - the rank stratum of the program in which the student ultimately enrolled. (*1*)  Setting aside for now the rank stratum, some such procedure, with the usual major caveats, would provide a conventional start to estimating which effects could be excluded from admissions decisions without causing major reductions in degree completion rates.

Several features of their analysis, however, contribute to major over-estimation of the statistical uncertainty in estimates of the predictive value of GREs, i.e. to the well-known "variance inflation" problem in estimating such parameters.(*7*) Inclusion of the rank stratum will also exacerbate *systematic* underestimation of the predictive power, already a problem due to lack of data on the students who were not admitted. (*8*) (*9*) (*10*) (*11*) The net result is to obscure the statistical reliability of the conclusion that those tests help predict which students are likely to get a PhD.

Variance Inflation from Collinearity

The main issue being addressed by the paper is not how well one can distinguish the separate predictive coefficients of GRE-Q and GRE-P but rather, since they show similar disparities among demographic groups (*1, 12*), what weight if any should be placed on such tests altogether. (*1*) (GRE-V turns out to have essentially no incremental predictive value.) The model shown includes both GRE-P and GRE-Q as separate variables. The scores on these exams are highly correlated(*2*), i.e. "collinear", which both divides up their net predictive power into two smaller pieces and inflates the uncertainties on the predictive coefficients of each variable (since the model fit is rather insensitive to their relative weights) (*13*). This can convert a highly significant net predictor into two that appear "insignificant". If, for example, one were to predict people's height from a model including right and left shoe sizes, the two shoe coefficients would be almost completely uncertain since the prediction doesn't care which variable is used. Either shoe could be dropped from the model. A very naïve reading of the statistical confidence ranges might suggest that neither shoe size was "significant" and therefore *both* shoes should be dropped from the model. Nevertheless the predictive coefficient for their *average* would be well-defined, and dropping both shoes from the predictive model would weaken predictive power substantially unless there were good substitutes. (*13*) The predictive coefficient for their sum would show relatively little statistical uncertainty.

To avoid the collinearity, one need only take the obvious step of combining GRE-P and GRE-Q into a single score, first scaling by the inverses of their ranges to make their contributions of equal weight, since unlike the shoes they have different ranges. (Their predictive powers are similar enough that almost nothing is gained by adding a parameter to give them unequal weights.) Although the follow-up paper(*2*) contains some general discussion of collinearity, the simple step of calculating the net GRE effect and its uncertainty for this obvious combination is omitted. The 10$^{th}$ to 90$^{th}$ percentile ranges for the U.S. group can be seen in Fig. 2 of (*1*), with GRE-P having ~1.5 times as large a range as GRE-Q in this cohort, so the equal-weighted sum is close to GRE-P+1.5*GRE-Q, i.e. 1.5*GRE-Q has about the same range as GRE-P. Using data from Table 2 of (*1*) its coefficient (GRE-P coefficient +(1/1.5)*GRE-Q coefficient) is virtually identical in the entire sample ("All Students") and the three subgroups described (U.S., U.S. female, and U.S. male): 0.0116 per percentile, within an accidentally small 1% range.

The paper(*1*) relies heavily on p-value significance cutoffs(*3*) to claim that the GREs are not predictors. The follow-up paper now gives the correlation coefficient between the estimated predictive coefficients for these two tests, -0.42. (*2*) Due to this large negative correlation the standard error in the estimated coefficient of the weighted sum, $SE_{SUM}=(SE_P^2+(SE_Q/1.5)^2-2*0.42* SE_P*SE_Q/1.5)^{1/2}$, is much less than it would be if the tests were independent. (Here I use the approximation that the other coefficients, e.g. for GPA, change little when the GREs are combined, although this combination has slightly different weighting than the one found with separately adjusted coefficients.) Using $SE_P$ and $SE_Q$ from Table 2 gives a standard error for the coefficient in the All Students group of 0.0026, less than ¼ of the coefficient's point estimate. In other words, based on the data of the paper and the follow-up, the GRE predictive effect is a 4.5-SE effect overall, far more statistically significant than any conventional cutoff value for such problems. In the three subgroups described (U.S., U.S. Male, U.S. female) it's a (3.4, 3.0, 1.5)-SE effect, respectively. (Each of these is approximate due to limited precision of the reported standard errors.) Each of these effect/SE ratios would increase further if the irrelevant GRE-V were dropped from the model. (Even simpler, just dropping either GRE-P or perhaps even GRE-Q would leave the other as clearly significant except in the small U.S. female subgroup, although not as strong as the simple equal-weight sum. Think of the shoes.)

To convert this robust coefficient to a net effect size for the combined GRE-P and GRE-Q we must allow for standard deviation (i.e. range) of the test combination being slightly less than the sum of the two separate standard deviations, which are equal for the range-weighted sum, by a factor of $((1+r_{PQ})/2)^{1/2}$ where $r_{PQ}$ is their correlation coefficient. Although no correlation data were included in the original paper, the subsequent note(*2*) gives $r_{PQ}$=0.55 for their sample, allowing us now to calculate the effect size for the sum. The 10$^{th}$ to 90$^{th}$ percentile effects for each test within the U.S. group are ~0.46 and ~0.36, respectively, reading the ranges and logit changes from Fig. 2 of (*1*), with slopes checked via Table 2 of (*1*). Using the approximation that the 10$^{th}$ to 90$^{th}$ percentile range for the sum scales like the standard deviation, the effect size from the 10$^{th}$ to 90$^{th}$ percentile is then a logit of $(1.55/2)^{1/2}(0.46+0.36)$ = ~0.72 in the U.S. subgroup. (Here I assume that slightly changing GRE weightings does not make a big change in coefficients of other variables.) Most of this predictive power could be obtained from the GRE-Q alone if GRE-P were dropped, but that would have little effect on the demographic disparities in the net criterion.

*The GRE effect is thus not only very statistically significant but also slightly larger than the effect size of GPA in the U. S. subgroup*, ~0.6 from Fig. 2 of (*1*). In the total sample, All Students, for which the predictive power of GPA apparently collapses by a factor of 2, (*1*) the GREs provide much greater predictive power than GPA. Thus, even before we get to the more interesting and serious systematic problems, we see that based on the data of Miller et al. (*1, 2*) the obvious weighted sum of GREs provides the best general predictor in their model in the group of all students and is approximately tied with GPA in the U.S. subgroup, although none of the predictors are very powerful. (Allowing different weights for GRE-P and GRE-Q, i.e. including their equal-weight difference as a predictor, adds very little to the predictive power.) Thus even before getting to the interesting and important modeling questions, we see that according to

the data of the paper(*1*) and addendum(*2*) overall GREs do better than GPA for predicting graduation within the context of the linear logit model.

Stratification: Variance Inflation, Confounding, and Collider-Like Bias

The model chosen includes the rank of the graduate program in which the student enrolled, via an adjustable extra term for three rank strata. (*1*) *Clearly this variable is not one that an admissions committee lacking pre-cognition could use to decide among competing applicants.* (Interaction terms between rank and other predictors could be used to help different programs choose different criteria, but no such terms are included in the model reported (*1*) or in the later addition.(*2*) ) Does rank nevertheless belong in a model estimating the predictive value of other metrics?

Before getting to the main problem from stratification- systematic bias in coefficient estimates- we see that it creates another variance inflation by restricting the range of the predictors. This problem of restricted range in predictive modeling is well known, especially in the context of educational and employment decisions (e.g. (*8*) (*14*)), and even in the specific context of physics GREs.(*9*) Correlation between outcomes and predictors is suppressed in narrow strata. In one experimental comparison correlation between scores on two components of a Swedish driving test fell by more than a factor of two when restricted to those who passed the first test.(*14*) In an especially relevant 1993 study(*8*), the GRE validity in predicting performance of psychology students in classes on statistics, assessments, and research methods was found to be high (0.55 to 0.70) in a program with little range restriction because GREs were not used in admissions decisions, in contrast to much lower validity in a range-restricted subset or to typical low validity for predicting grades in more range-restricted programs which use GREs for admissions decisions. The authors' conclusion was "These results support the conventional argument that uncorrected GRE validity estimates based on range-restricted samples are strongly biased toward zero." (*8*)

What is unusual about the Miller et al. analysis is not that there was a restricted range problem, since a school or employer typically does not have performance data on the those who either were not offered a position in their institution or did not choose to take it. What's unusual is that the range restriction here was largely self-inflicted by stratifying the students by program rank. (*1*) Miller et al. state that one of the strengths of their study is that it includes a wide range for the predictive variables because it includes schools of very different ranks, (*1*) but they do not fully use that range to narrow the statistical uncertainties in the parameter estimates.

Is that loss of precision justified by the need to avoid systematic errors? Miller et al. say they "…include covariates to render more precise estimates", but including covariates can either remove or add systematic bias terms depending on which covariates are included and on what one wishes to estimate.(*6*) (*10*) (*11*) In causal inference studies stratifying on a "collider", a downstream variable affected both by the suspected cause and by unmeasured other causes, adds a new systematic error, called collider stratification selection bias, to the causal estimand.(*6*) (*10*) (*11*). In one famous case, inadvertent collider conditioning gives the paradoxical effect that maternal smoking appears to protect low birth weight newborns from mortality, because within the low birth weight stratum smoking is negatively correlated with even more ominous predictors. (*15*)

Miller et al. find that even after taking into account GPA, GREs, etc. students in the higher-ranked programs have a higher likelihood of completion. (*1*) Using their stratified model to evaluate the incremental predictive power of GREs implicitly assumes that this boost is caused entirely by factors that would not change if students with lower scores were admitted to those programs. There are two main possible causes of this boost mentioned in the original paper.

One possibility mentioned(*1*) would be that a typical student has a systematically easier time graduating from higher-rank programs than from lower-rank programs, so the boost would persist even if admissions procedures changed and students who would currently enroll in

6/7/20    8

lower-rank programs were switched to high-rank programs. If this were the main explanation, then rank would be a simple confounder and should be removed by stratification or other methods to improve the estimate of the incremental predictive power of the GREs. No evidence is given to support this possibility, and the actual sign of the effect is not obvious. We shall see that internal evidence of the paper tends to contradict this explanation.

The other possibility is that the high-ranked programs are getting students with a higher propensity to graduate than predicted by the in-model GREs and GPA because they use a variety of other predictors as well, as has been thoroughly documented(*16*) by a group sharing an author with Miller et al.(*1*). These predictors include prior research experience, letters of recommendation, etc. (*16*). Unless that ubiquitous effort is pointless, these predictors will have some positive predictive value, which will be reflected in the coefficient of the rank variable, with which they will be positively correlated. (*1*) If the out-of-model predictors are positively correlated with an in-model predictor, they will increase the coefficient that the model assigns to that predictor beyond what would actually be lost by dropping the predictor, but if they are negatively correlated they will decrease that coefficient. The model estimate will depend on stratification because the correlation between the in-model and out-of-model predictors changes as a function of stratification. (*10*)

Students with low GREs and GPAs who nonetheless are accepted into high-rank schools are likely to have especially good prior research experience, letters of recommendation, etc., creating a *negative* correlation within each stratum between those routinely used out-of-model predictors and the predictors used in the model. (*9*) In an analogous case, although performances on long-jumps and 110 meter races are likely to be positively correlated in the general population, in the stratum of Olympic decathletes these have a strongly negative correlation. (*17*)

The reported data include indications that the odds boost for students in high-ranked programs is likely to be due primarily to the out-of-model predictors used in admissions rather than to any direct student-independent effects (of unknown sign) of differently ranked programs. If some randomly chosen students were boosted in enrolled program rank, their graduation probability would increase from the hypothetical direct effect but not change for the out-of-model selection effect. In the selection case, but not the direct effect case, the stratified model would then assign this random group a negative logit equal to the positive logit assigned to the rank boost. In a causal diagram, the random group assignment would collide with effects of out-of-model selection traits on program rank, and the random group assignment would pick up a logit via collider bias despite having no causal effect on graduation. Something approximately similar to that randomized trial would happen if the boosted students were picked non-randomly, but based on traits with little direct relevance to graduation probability. Given the almost universal attempt to boost representation of under-represented minorities, we may see such statistical artifacts in the large negative logits the model assigns to them (seen in Table 2 (*1*)), statistically significant in the overall sample and close in magnitude to the positive logit assigned to the difference between the first and third rank tier. That pattern is more consistent with collider bias in the model than with the more selective programs being easier to complete, although without further information on other possible factors one cannot precisely sort out such systematic effects. I predict that these negative demographic logits will shrink substantially in a less-stratified (and, as I will argue, probably more accurate) model omitting program rank, and could easily fall to zero or turn positive if a fully unstratified model or one including all important predictors were possible. It is ironic that the same stratification collider bias that helps minimize the model's estimate of the predictive value of GREs, nominally for the sake of under-represented minorities, produces as collateral damage negative predictive logits for those groups, even after controlling for the effects of GREs and GPA.

Since the out-of-model predictors are themselves likely to be positively correlated with in-model predictors, they would be confounders in a model completely lacking range restriction, causing some positive overestimate of the incremental predictive power of the in-model predictors. For the real data, however, the unavoidable limitation to students who have been accepted means that the population under study is systematically restricted compared to the one of interest- all the applicants plus some others who might apply if GREs were dropped. (*9*)
That unavoidable range restriction effect is not small. For example, if both in and out
contributions are independent normally distributed and given equal weight, mere selection of
applicants with an overall above-average score gives a correlation coefficient of
$-1/(\pi-1) = -0.47$. Even if the in and out predictors are positively correlated (coefficient $r_{OI}$) in the
entire applicant population, their correlation in the enrolled upper half is
$((\pi-1)r_{OI}-1)/(\pi-1-r_{OI})$. Even without rank strata, the model would underestimate the in-model
coefficients if $r_{OI} < 1/(\pi-1) = 0.47$, which is larger than one would ordinarily expect the
correlation to be between disparate predictors such as test scores and research experience.
Since the coefficients of the tiers do not show especially large variance inflation(*2*), they cannot
be very strongly correlated with the other predictors. (Obviously it would be far easier to
reason briefly and accurately about this if the covariances between program tier and other
variables were given.) Thus to the extent that the positive logits for high-ranked programs are
caused by their selection of students, even a model omitting rank strata would be likely to
underestimate the incremental predictive power of including GREs, or at any rate not
overestimate it by very much.

The more finely rank is stratified, the more negative these correlations become. (*9*) *In the ideal limit of narrow rank stratification and admissions criteria successfully aimed to maximize a particular goal, all power for predicting that goal using any variables other than rank becomes zero regardless of how predictive they are in the unstratified population, since no variation is left within each stratum.* That remains true regardless of how much range remains for any individual predictor, how complete the overall range of the data is, and how large the sample size is. That program rank should be a relatively good predictor in the highly stratified model thus tells us little other than that admissions committees are making use of the out-of-model predictors that they say they use. (*16*)

The follow-up paper (*2*) gives some more range data, but entirely ignores the stratification bias issue and does not give the ranges of scores in the different strata. The arguments presented assume that mere inclusion of the different strata would get rid of problems, rather than recognizing that the collider stratification itself creates systematically biased estimates of predictive coefficients.

Null Hypotheses for Subsamples, Anomalous Confidence Intervals, Dynamic Range Compression

The Miller et al. paper reasonably avoids making a strong prior assumption that each predictor will work equally well in each subsample. As we've seen, however, the point estimate for the net GRE predictive coefficient based on their data is virtually identical in each subsample, providing no evidence that net GRE weighting should differ among them. The paper replaces the conventional null hypothesis of equal effects in different subsamples with null hypotheses of no effect in each subsample. This gives anomalous interpretations. For example, although the point estimate given in Table 2 for the coefficient of the logit for GRE-Q in All Students (0.013 per percentile rank) is statistically significant, and the point estimate among U.S. females (0.017) is larger, the latter fact is described as "we see no differences in Ph.D. completion probability…" in females. (*1*)  Here, as elsewhere (including the quotes in my first paragraph) the paper makes the common misinterpretation that results insufficiently precise to confidently reject a null hypothesis are equivalent to confirmation of the null.(*3*) That misinterpretation often leads to confusion and "confirmation" of contradictory hypotheses. (*3*) In typical medical trials when a treatment appears to work better in a subsample than in the overall group, but with larger uncertainty due to the small sample, one doesn't conclude that the treatment doesn't work in the smaller group, even though that possibility cannot be ruled out.  The treatment of the null used by Miller et al. (*1*) would routinely lead to conclusions such as that although a treatment worked well overall it would not work at all in *any* particular group of people, since the uncertainties in any small group are large.

Figure 2 of (*1*) illustrates the predictive slopes of the U.S. subsample for GPA and the GREs applied separately to 10[th], 50[th], and 90[th] percentile scores for U.S. females and males.  It shows very large "95% confidence intervals associated with Ph.D. completion probability", leaving the visual impression that predictive effects are small compared to uncertainty. Converting  to logits, these intervals are ~± 1.1 for each estimate at the low, middle, and high parts of the distributions for both U.S. males and females. The near-equality at the middle and edges of the distribution tells us that these intervals cannot primarily reflect the uncertainty of interest, i.e. uncertainty in the slopes of the logit dependence on the model variables, because that would not show up much in the middle points. For large N in the middle of the parameter range the 95% confidence intervals for the logit should be $\pm 1.96*/(Np(1-p))^{1/2}$. For the full U.S. sample with N=2315 and p =~0.7, that would be ±0.09, not ±1.1. The confidence intervals shown appear to be based on the number of students (~23) within each integer percentile group rather than the actual group size from which the probability estimates are calculated, and thus are inflated by approximately one order of magnitude.

It may strain credulity to claim that anyone would use such an integer-percentile procedure to estimate confidence intervals in the context of a linear logit model for which the point estimates are based on all the data, but something like that seems to have been done. The visual effect of these radically inflated confidence intervals is to de-emphasize the predictive power of the admissions criteria even beyond the substantial variance inflation introduced by the model itself.

Rather than directly use the GRE scores themselves in the linear model, the paper uses percentile rankings(*1*), a convenient way to stitch together scores from before and after the GRE scale changed. It is not required, however, since score conversion tables are available. The percentile method has the effect of greatly compressing the dynamic range in the higher scores in the tail of the distribution and magnifying small differences in the meat of the distribution, where most accepted applicants are found. Thus it is possible that this highly non-linear map from test scores to the predictors used in the linear model reduces the predictive power.

Some smaller features of the presentation style are also problematic. For predictors whose power the authors wish to emphasize (e.g. program rank) the results are often presented in terms of odds ratios. For those whose predictive power the authors wish to deemphasize (GREs) the results are always presented in terms of percentage differences in completion rates. A comparison of completion rates of 75% and 60% gives a rather small-sounding 15% rate difference, a medium-sounding logit of 0.69, and an odds ratio of 2, which sounds rather large.

Key data (the results of a model without rank stratum, information on covariance of stratum with other predictors, ranges of variables on the overall group, predictive coefficients for the large non-U.S. group, etc.) remain missing even in the belated follow-up release.(*1, 2*) Since this missing information does not appear in the arXiv follow-up(*1, 2*), it cannot have been omitted simply due to space limitations of the original paper. A small anomaly appears in Table 2 for the "non-U.S." group, whose group logit is given as positive 0.09 but whose group odds ratio is given as 0.9, i.e. $e^{-0.09}$ rather than $e^{+0.09}$. (*1*)

<u>The Bottom Line</u>

Based even on the incomplete data presented, the *statistical* uncertainty in estimating how much predictive strength would be lost by dropping or de-emphasizing GREs is not particularly important. (*1*)   We've seen that in the U.S. subsample a simple equal-weight sum of the two relevant GREs provides a logit difference of ~0.72, i.e. an odds ratio of ~2.1, even before making any upward correction for the systematic stratification bias or for possible improvement from using test scores rather than percentiles.

Extending those results to the non-U.S. 40% of the sample requires guesswork, because the range data and correlation coefficients for that subsample have not been provided. In the published results, there is no indication that GREs would be a weaker predictor in that group than in the U.S. (*1*)  In contrast, the predictive coefficient for GPA is only about half as big in All Students as in the U.S. (*1*)  Thus although no predictors of graduation are especially good, the net equal-weight GRE-P&Q combination looks better than GPA overall.

Results in the addendum(*2*) for formal model evaluation criteria, which include a likelihood measure and a penalty for adding parameters, look consistent with this conclusion, although a simple model including GPA and the net GRE-P&Q but omitting the irrelevant GRE-V(*1*) is not included. It is not hard to extrapolate from the models given to see that by the standard Akaike Information Criterion (AIC) this simple model would be superior to any of them by very substantial margins in the All and U.S. groups, significantly in the U.S. male subgroup, and approximately tied with GPA alone in the U.S. female subgroup. Extending the results to lower scores, particularly relevant for GRE-Q whose range is strongly restricted in the sample, not just the strata, (*2*), is uncertain, but past indications are that such dependences do not become any weaker in the low end. (*8*)

Due to the collider bias from stratification of the model on admissions and post-admissions program rank all these logit differences, both for GREs and for GPA, probably substantially underestimate the predictive power that would be lost to admissions committees by dropping the predictors in the model. Use of percentiles rather than scores may have also led to underestimation for the GREs. There is no reason to believe that the slope of logit vs. GRE percentile would become weaker if the range of GREs accepted were extended downward. (*8*)

One caveat is in order. These data give no compelling reason to think that fields other than physics, or even various subfields in physics, will find similar predictive effects. It's quite possible that, e.g., GRE-Q is more predictive for physics and for quantitative social science methods(*8*) than for many other fields.

Discussion

Miller et al. have collected a substantial amount of useful data on a question of wide interest. Since their data set includes a relatively large predictor range it opens up the possibility of making more reliable estimates than ones based on data or anecdotes from individual programs. Some of their results are already useful, e.g. showing the GRE-V adds essentially nothing to the predictive power in this sample. (*1*) The correlation coefficients of the addendum(*2*) let us see that an equal-weight GRE-P&Q combination is only a slightly stronger predictor than plain GRE-Q, a result which may have little import for admissions decisions but might matter in evaluating costs of admission requirements. (There are anecdotal reports that the direct financial cost of an added test is a barrier for many students, especially from under-represented groups.) There is also a slight statistical hint, about a 1-S.E. effect, suggesting that uequal weights for GRE-Q and GRE-P might be useful for U.S. females.

Unfortunately, the analysis employed in the original paper systematically obscures some key implications of the data. The vulnerability of sloppy statistical arguments to confirmation bias is not new. In particular, the problem of "p-hacking" or "data-dredging" is well-known. (*18*) Motivated researchers can search among many hypotheses to find ones that happen by accident to meet the arbitrary conventional p-value criterion for "significance" (*3*).

The Miller et al. paper (*1*) appears to be an instance of reverse p-hacking.(*19*) Some "insignificant" p-values are sought and found to confirm the lead author's often-repeated claim (e.g. (*12*) (*20*)) that "the US Ph.D. completion rate in STEM fields is only 50%…. So the standard admissions procedure is no better a predictor of success than a coin flip."(*20*) The logic of that claim is identical to that of a claim "Since the 5-year survival rate is only 50% the treatment is no better than a placebo", as if the expected outcomes for the untreated condition, e.g. pancreatic cancer or acne, were irrelevant. Of the athletes admitted to the U.S. Olympic track trials, less than 10% graduate to the Olympic team. Is the selection procedure for Olympic trial athletes a worse predictor of success than pure chance would be?

Finding spurious negative results is even easier than finding spurious positive results, especially when one is free to search through a variety of models before choosing one for "parsimony".(*1*) One need only combine a few variance-inflators and some stratification on downstream variables with a willingness to misinterpret failure to reject the null on some subsamples as confirmation of the null. (*3*) The claimed null value of the GREs as predictors is an artifact of these improper procedures.

The methodological problems of the paper are a reminder of why some social science journals are instituting a "registered report" system, similar to the protocols now used for medical trials. For approved medical treatment trials, since experimenters (e.g. drug companies) typically have intense motivation to find some positive results, they are required to file protocols ahead of time specifying which outcomes will be tested by which statistical techniques. A similar "registered report" system is now spreading to social sciences. (*21*) That approach might be especially valuable for studies directly related to policy choices, analogous to medical treatment choices.

Only a few straightforward further steps would be needed to make much fuller, more transparent use of these data. Most important by far would be simply to publish the results for the model without the biasing stratification by program rank. Another very useful step would be to publish results (ranges of predictors, coefficients, and correlations ) on the 40% of the sample that are not U.S. citizens, just like on the U.S. subsamples. Publishing ranges of variables for "All Students" would also be useful. Publishing a comparison of a model which drops GRE-P with one that uses an equal-weight P&Q sum would help in cost-benefit decisions about requiring GRE-P. All of these steps would require virtually no work. Release of such summary analyses would seem to be required by the AAAS publications policy: "After publication, all data and materials necessary to understand, assess, and extend the conclusions of the manuscript must be available to any reader of a Science Journal. After publication, all reasonable requests for data, code, or materials must be fulfilled." A small amount of additional calculation would also allow an analysis using scores rather than percentiles.

Even a transparent analysis of these data will potentially be subject to systematic errors of unknown sign, as with any observational study. Since there are now fairly many physics departments unsure of what GRE policy to adopt, a randomized controlled trial now might directly address the causal policy question: what effects do different admissions policies have on student outcomes? Different volunteer departments could be randomly assigned to different GRE policies for a year, then switched for the next year. The resulting cohorts could then be followed not only for degree completion but for other more important outcomes. Most importantly, regardless of the results, it would set an example of scientists trying to use objective scientific methods to help make policy choices. Such an example might be particularly valuable in the current climate in which distrust of even the best-founded and most important scientific claims on the part of a substantial fraction of the population can have disastrous effects.

Reply to the Authors' Response

The original authors have published a nominal response(*5*) to my published critique.(*4*) It essentially doubles down on the key errors.

They again decline to present the results for a simple model like the one they initially used but without the biasing rank stratification. They again decline to report the correlation coefficients between rank and the other predictive variables. (*5*) In response to the point that stratification by program rank negatively biases coefficient estimates and forces those coefficients to zero in an ideal limiting high-stratification case (*4*), they replaced rank with *individual program variables*, i.e. *increasing* the stratification. (*5*) In response to the point that aggregating stratification-biased results does nothing to remove the systematic bias(*4*), they again aggregated the even-more-stratification-biased results, presenting "mean-centered" results. (*5*)

In response to my calculation of the actual predictive effective of a combined GRE score, based on the actual within-sample correlation between GRE-P and GRE-Q from their own report(*4*), they present a series of qualitative arguments about why it shouldn't be necessary to do such a calculation. (*5*)  They simply do not address the actual result, that the net GRE score is somewhat more predictive than GPA even in the U.S. subsample, and substantially more overall. (*4*)

In response to the point that the significance of GRE-Q would be evident even in their model if they just dropped the substantially collinear (and somewhat less predictive) GRE-P(*4*), they presented only results for GRE-P, dropping GRE-Q rather than GRE-P! (*5*) Furthermore, even though these results used only the less predictive GRE-P and were now ultra-stratified by individual program, the coefficient of GRE-P barely missed nominal significance at p=0.06. (*5*)

In response to the point that they had omitted the obvious model from their AICc model evaluations in the addendum(*4*), they repeated the results from the addendum(*2*) without including the obvious model. (*5*) Most importantly, they do not compare a model without the GRE-P +GRE-Q sum to one including that sum, so that one cannot directly see how including that sum contributes to the AICc.

The do not respond to the point that the negative logits reported for under-represented minorities in their initial analysis (*1*) indicated substantial negative collider bias(*4*). They do not report those logits for the even more highly stratified model used in the response(*5*).

As justification for the use of a highly stratified model, which they now concede can introduce bias, they argue that programs of different types may have different needs. (*5*) While true, the mathematical expression of such differences takes the form of interaction terms between the predictors of interest (e.g. GRE-Q) and the program rank or (in the new version) individual program variables. Such interaction terms were not included in the previous reports (*1*) (*2*), so that the reported results could not be used to obtain different criteria for different programs. The current report now says that such interaction terms were insignificant (*5*), thus again negating the stated rationale for including the biasing rank or program covariates.

In response to the point that the confidence intervals presented in Fig. 2 of (*1*) cannot represent the uncertainty in estimating the coefficients of interest and are too large by an order of magnitude(*4*), they include some discussion of these confidence intervals. It is impossible to evaluate that discussion since they still decline to define what those confidence intervals are intended to represent, or what relevance they have to the coefficients of interest. (*5*) In particular, they do not address the question of whether they calculated the confidence intervals for graduation probability for a median student as if it were estimated based on only 1% of the sample, which is what they seem to have done. (*4*)

In response to the point) that the effects of using GRE-aware or GRE-blind admissions could better be assessed via a randomized trial(*4*), they argue that causal estimations from randomized trials are subject to "non-trivial biases, including collider effects, due to programs' voluntary participation." (*5*) This is qualitatively incorrect because the point of randomizing treatment is to sever all back-door paths, eliminating classical confounding. Collider effects would enter only through non-random *post-randomization* drop-outs, not initial pre-randomization volunteering. This issue is routinely dealt with via intent-to-treat analysis. (*6*) (*10*) (*11*) Therefore the restriction to voluntary participant departments would not produce causal estimand bias although, as in any such trial, it could limit the generalizability of the results. The response also argues that even the limited resources involved in evaluating the predictive effectiveness of GREs by randomly assigning policies to programs unsure about what policy to use would waste resources that should instead be used to implement the policies they advocate. It is unclear whether this argument is consistent with the rationale for their multi-year study on the predictive effectiveness of GREs, funded by four NSF grants. (*1*)

In discussing range restriction, they raise various points that are irrelevant to the issue of systematic bias. (*5*) They concede that their analysis, necessarily confined to those currently admitted to grad school, is not intended to apply "for the entire population of test takers". (*5*) The point of the study was to see if changing admissions procedures to admit students who are now excluded because they either scored very low or even declined to take the exams would result in a lower graduation rate. Therefore the ability to extrapolate toward the lower end is actually crucial for practical application of the results, even if they were analyzed correctly.

In response to the point that using GRE percentiles rather than scores might reduce the predictive power(*4*), they present some plots of the distributions, Fig. 3 of (*5*), that do suggest that any such effect is not huge. (*5*) Nonetheless, it would be useful to see the actual results obtained with scores.

They do not respond to the point that they have presented no results broken out for the 40% of the students who are not U.S. citizens(*4*), in contrast to the results for U.S. citizens. (*1*) (*5*) Perhaps their objection that "these scores can be efficiently used to eliminate … U.S. citizens from the discipline" (*5*) accounts for omitting analysis of the criteria for non-U.S. citizens. On the other hand, given the exclusionary policies of the current federal administration, perhaps the authors were prescient in seeing that this point would become moot.

In summary, the response (*5*) intensifies rather than removes stratification bias and then cherry-picks the weaker predictor (GRE-P) rather than the somewhat stronger one (GRE-Q) or the strongest one (their equal-weight sum) to analyze. It thus clarifies that reverse p-hacking(*19*) was indeed employed.

Conclusion

Even a randomized trial would not entirely determine the effects of changing criteria, which may not even be dominated by the individual-level effects discussed here. There are many ways in which the Stable Unit Treatment Value Assumption(*22*) would fail to apply, complicating any estimate of the net effects of policy changes. Changes in institutional traits and motivational signaling effects would be particularly hard to predict. For example, if many prominent graduate departments were to drop consideration of GRE-P in admissions decisions, many institutions might change undergraduate physics curricula and grading standards, for better or worse or both.

Meanwhile, the question remains of what use should be made now of the actual predictive power of the GREs. That involves non-technical considerations rather than p-values. The issue of how our profession should choose its new members faces a variety of not always parallel social goals and is fraught with uncertainties. Despite these difficulties, ***finding the best selection method is trivial in one limiting case. If we do not try to maintain minimal standards of competence and transparency or even basic logic in our treatment of data, then the optimum group of students whom we should be educating is the empty set.***

Acknowledgements: I thank Ellen Fireman, Jamie Robins, Alex Small, and many others for extremely helpful conversations, but none of them are responsible for the contents of this paper. I requested information on what I thought were discrepancies between Table 2 and Figure 2 of (*1*) and on the missing covariance matrix from the corresponding author on 2/12/2019 and again on 2/23/2019 and from the other authors on 2/24/2019. Although the paper says "Additional data relating to this paper may be requested from the authors" on 2/25/2019 the corresponding author wrote that issues involving "human subjects" would need to be resolved before these summary statistics could be shared. On 6/20/2019 the corresponding author sent a first draft of the follow-up paper, containing some parts of the covariance matrices, for which I thank him.